\documentclass{emulateapj}
\usepackage{apjfonts}

\newcommand{\ltsimeq}{\raisebox{-0.6ex}{$\,\stackrel 
        {\raisebox{-.2ex}{$\textstyle <$}}{\sim}\,$}} 
\newcommand{\gtsimeq}{\raisebox{-0.6ex}{$\,\stackrel 
        {\raisebox{-.2ex}{$\textstyle >$}}{\sim}\,$}}

\newcommand{\lya}{Ly\,$\alpha$}
\newcommand{\lyb}{Ly\,$\beta$}
\newcommand{\lyg}{Ly\,$\gamma$}

\newcommand{\asec}{^{\prime\prime}}

\def\hst{{\it Hubble Space Telescope~}}

\shorttitle{First results from the Canada-France High-$z$ Quasar Survey}
\shortauthors{Willott et al.}

\begin{document}


\title{First results from the Canada-France High-$z$ Quasar Survey:
Constraints on the $z=6$ quasar luminosity function and the
quasar contribution to reionization}



\author{
Chris J. Willott\altaffilmark{1},
Xavier Delfosse\altaffilmark{2},
Thierry Forveille\altaffilmark{3,2},
Philippe Delorme\altaffilmark{2},
Stephen D. J. Gwyn\altaffilmark{4}
}

\altaffiltext{1}{Herzberg Institute of Astrophysics, National Research
Council, 5071 West Saanich Rd, Victoria, BC V9E 2E7, Canada;
chris.willott@nrc.ca} 
\altaffiltext{2}{Laboratoire d'Astrophysique de Grenoble, BP 53, 38041
Grenoble Cedex, France; Xavier.Delfosse@obs.ujf-grenoble.fr,
Philippe.Delorme@obs.ujf-grenoble.fr }
\altaffiltext{3}{Canada-France-Hawaii Telescope Corporation, PO Box
1597, Kamuela, HI 96743, USA; Thierry.Forveille@cfht.hawaii.edu}
\altaffiltext{4}{Dept. of Physics \& Astronomy, University of
Victoria, PO Box 3055, Victoria, BC V8W 3P6, Canada;
gwyn@uvastro.phys.uvic.ca}

\begin{abstract}
We present preliminary results of a new quasar survey being undertaken
with multi-colour optical imaging from the Canada-France-Hawaii
Telescope. The current data consists of 3.83 sq. deg. of imaging in
the $i'$ and $z'$ filters to a $10\,\sigma$ limit of $z'<23.35$.
Near-infrared photometry of 24 candidate $5.7<z<6.4$ quasars confirms
them all to be low mass stars including two T dwarfs and four or five
L dwarfs. Photometric estimates of the spectral type of the two T
dwarfs are T3 and T6. We use the lack of high-redshift quasars in this
survey volume to constrain the $z=6$ quasar luminosity function. For
reasonable values of the break absolute magnitude $M^*_{1450}$ and
faint-end slope $\alpha$, we determine that the bright-end slope
$\beta>-3.2$ at 95\% confidence. We find that the comoving
space-density of quasars brighter than $M_{1450}=-23.5$ declines by a
factor $>25$ from $z=2$ to $z=6$, mirroring the decline observed for
high-luminosity quasars.  We consider the contribution of the quasar
population to the ionizing photon density at $z=6$ and the
implications for reionization. We show that the current constraints on
the quasar population give an ionizing photon density $\ll 30\%$ that
of the star-forming galaxy population. We conclude that active
galactic nuclei make a negligible contribution to the reionization of
hydrogen at $z\sim 6$.

\end{abstract}


\section{Introduction}

After the combination of protons and electrons at redshift $z\sim
1000$ the hydrogen content of the universe remained largely neutral
for some time. Observations of high-redshift quasars show that the
neutral hydrogen in the diffuse intergalactic medium (IGM) became
fully reionized by $z\approx6$ (Fan et al. 2004). In the standard
picture (e.g. Barkana \& Loeb 2001) this reionization was the result
of a rapid rise in the production of hydrogen-ionizing photons due to
the birth of the first sizable population of stars and accreting
black holes.

A census of the star-forming galaxy population at $z\approx 6$ is now
possible due to deep optical surveys carried out by the \hst\
(Dickinson et al 2004; Bunker et al. 2004; Bouwens et
al. 2004). Whilst the luminosity function of these galaxies is now
fairly well constrained (Bunker et al. 2004), translating this to the
UV ionizing photon production rate relies upon a couple of uncertain
factors: the spectral shape shortward of \lya\ (which depends upon the
stellar initial mass function and metallicity) and the fraction of
ionizing photons which escape the host galaxy. Therefore the ionizing
photon production rate of the star-forming galaxy population remains
quite uncertain and it is not clear whether this rate is high enough
to reionize the universe at $z=6$ (Stiavelli, Fall \& Panagia 2004).

An alternative source of ionizing photons is the quasar population.
Quasars have very hard spectra which makes them extremely efficient
ionizing sources. It is possible that the earliest phases of
reionization at $z>10$ were carried out by low-mass accreting black
holes also known as `mini-quasars' (Madau et al. 2004). The $z=6$
quasar luminosity function is much more poorly constrained than the
$z=6$ galaxy luminosity function and consequently the quasar ionizing
photon output is very poorly known. The only quasars known at this
redshift are from the bright Sloan Digital Sky Survey (Fan et al. 2004
and refs therein) and these luminous quasars make up only a tiny
fraction of the total quasar luminosity density. The bulk of the
luminosity density is emitted by quasars close to the `break' in the
luminosity function.

Attempts to constrain the high-redshift quasar ionizing photon output
have therefore largely depended upon extrapolation of the SDSS quasar
space density to lower luminosities using parameters derived for the
luminosity function at lower redshifts (Fan et al. 2001; Yan \&
Windhorst 2004; Meiksin 2005). This necessarily involves large and
uncertain extrapolations. These studies show that, for the {\em
expected} values of the luminosity function parameters, quasars fall
short of producing enough ionizing photons for reionization. However,
there is considerable debate as to how many photons are necessary for
reionization since there is a large uncertainty in the clumpiness of
the IGM which controls the importance of recombinations (Madau, Haardt
\& Rees 1999; Haiman, Abel \& Madau 2001; Oh \& Haiman 2003; Iliev,
Shapiro \& Raga 2005; Meiksin 2005).

Dijkstra, Haiman \& Loeb (2004) presented a novel method to constrain
the quasar contribution to reionization by considering the unresolved
component of the soft X-ray background. They found that, for their
adopted quasar spectral shape between the UV and hard X-rays and
clumpiness of the IGM, $z=6$ quasars cannot provide enough photons to
reionize the universe at $2\sigma$ significance. Meiksin (2005)
discussed this soft X-ray background constraint and showed that
adjusting either the typical quasar spectral shape or clumping factor
within reasonable ranges produced a result consistent with
reionization at $z=6$ by quasars.

Current constraints on the $z=6$ quasar luminosity function also
depend heavily on the properties of SDSS quasars. The luminosity
distribution has been used to place a $2\sigma$ range on the
bright-end slope of $-4.2<\beta<-2.2$ (Fan et al. 2004). The lack of
strong gravitational lensing amongst the SDSS sample can also place a
constraint on $\beta$ (e.g. Wyithe \& Loeb 2002; Comerford, Haiman \&
Schaye 2002).  Fan et al. (2003) and Richards et al. (2004) showed
that this leads to a constraint in the range $\beta>-3.5$ to
$\beta>-4.5$ at $2\sigma$ significance depending upon the assumed
values for the break magnitude and faint-end slope. Wyithe (2004)
combined SDSS lensing constraints and the luminosity distribution of
Fan et al. (2003) to find that $\beta>-3.0$ at 90\%
confidence. However, using the steeper slope preferred by the
luminosity distribution of the expanded sample of Fan et al. (2004)
would have the effect of weakening the conclusion of Wyithe (2004). A
search for lower luminosity quasars at $z>5$ in the Chandra Deep
Field-North found one quasar at $z=5.2$ but none at higher redshift
(Barger et al. 2003).

The Canada-France High-$z$ Quasar Survey (CFHQS) utilises the
wide-area deep surveys now being undertaken at the 3.6m
Canada-France-Hawaii Telescope with the 1 degree field-of-view camera
MegaCam. In this paper we present the method for selecting $z>5.7$
quasars and our initial results. The data used in this paper represent
a very small fraction of the full area/depth that is planned.

The paper is organised as follows. In Sec.\,2 we discuss the expected
optical/near-IR colours of point sources (quasars and stars) as
observed with MegaCam and the near-IR camera CFHT-IR. In Sec.\,3 we
describe the optical and near-IR datasets and the search for
high-redshift quasars.  In Sec.\,4 we discuss the limits on the
high-redshift quasar luminosity function which result from our
work. Sec.\,5 addresses whether $z\approx 6$ quasars can produce
enough UV photons to reionize the universe. Cosmological parameters of
$H_0=70\,{\rm km\,s^{-1}\,Mpc^{-1}}$, $\Omega_{\mathrm M}=0.3$ and
$\Omega_\Lambda=0.7$ are assumed throughout.

\section{Simulating the colours of quasars and stars}
\label{simul}

\subsection{High-redshift quasars}
\label{simulq}

The sharp drop in flux across the \lya\ emission line in high redshift
quasars makes colour selection the most efficient method for selecting
quasars over large areas (e.g. Richards et al. 2002). This technique
has been successfully utilised in the SDSS to discover all of the
highest redshift quasars currently known (Fan et al. 2004).  Spatially
unresolved sources with $i'-z'>2.2$ in the SDSS system are either
quasars at $z>5.7$ or brown dwarfs. The filter transmission profiles
and CCD sensitivity for MegaCam are not identical to those of the
SDSS. Therefore to determine the optimum colour selection criteria for
high-redshift quasars with CFHT, we have performed simulations of the
expected colours of quasars in the CFHT filters.

Previous methods to simulate the colours of high-redshift quasars have
used composite quasar spectra or power-law + emission line spectra as
the input quasar template. To obtain a realistic spread in quasar
properties, we use the observed spectra of a sample of $z\approx 3$
quasars to simulate the colours of quasars at higher redshift. A
similar method has recently been presented by Chiu et al. (2005).
However, their analysis was for the SDSS filters and differed from
ours in several aspects, so we outline our method in detail
here. 

Since our high redshift quasar selection is based on the $i', z'$ and
$J$ filters, we require optical quasar spectra which sample the
rest-frame wavelengths of $5.5<z<6.7$ quasars in these filters. This
wavelength range can be achieved with the SDSS DR3 spectroscopy of
quasars at $3.1<z<3.2$. At these redshifts, there is very little
colour incompleteness in the SDSS, due to the fact that the quasar
colours lie far from the stellar locus (Richards et al. 2002). This is
important since we are using these quasars to assess the efficiency of
our own colour selection criteria. There are 180 quasars in the DR3
catalogue satisfying these criteria. The lack of evolution in quasar
spectra from $z=6$ to lower redshift (Fan et al. 2004) justifies our
use of lower redshift templates. We assume the range of emission line
strengths, broad absorption line occurrence and spectral shapes of the
$z=3$ quasars are representative of the $z=6$ quasar population.

The only significant difference between the rest-frame UV spectra of
$z=3$ and $z=6$ quasars is the much stronger HI absorption at higher
redshift. Therefore the spectra of the $z=3$ quasars need to have
their Lyman forest absorption properties changed to be consistent with
being located at a higher redshift. The Lyman forest absorption over
the redshift range $2<z<6$ has recently been studied using
high-resolution spectroscopy by Songaila (2004). The mean transmission
for \lya\ as a function of redshift is taken from their eqn. 3. This
relation only applies at $z>4$, so the lower redshift transmission was
determined by fitting to the data tabulated in that paper. The \lyb\
optical depth was determined using $\tau_\beta=0.40\tau_\alpha$ as
found by Songaila.  First of all it is necessary to statistically
correct for the \lya\ and \lyb\ absorption already present in the
spectra of the $z=3$ quasars. \lyg\ and higher order absorption are
not considered since they have a very small effect on the flux
integrated over any of our filters at any redshift. We also set all
flux below the quasar Lyman limit to zero. After correcting for the
lower redshift absorption, absorption at higher redshifts assuming the
quasar is located at $5.5<z<6.7$ is applied. For each $z=3$ quasar, we
introduce scatter in the Lyman forest absorption as a function of
redshift (assuming the quasar lies at $5.5<z<6.7$). The amount of
scatter in $\delta z =0.1$ bins is estimated as a function of redshift
from the observations of Songaila (2004).

\begin{figure}
\resizebox{0.48\textwidth}{!}{\includegraphics{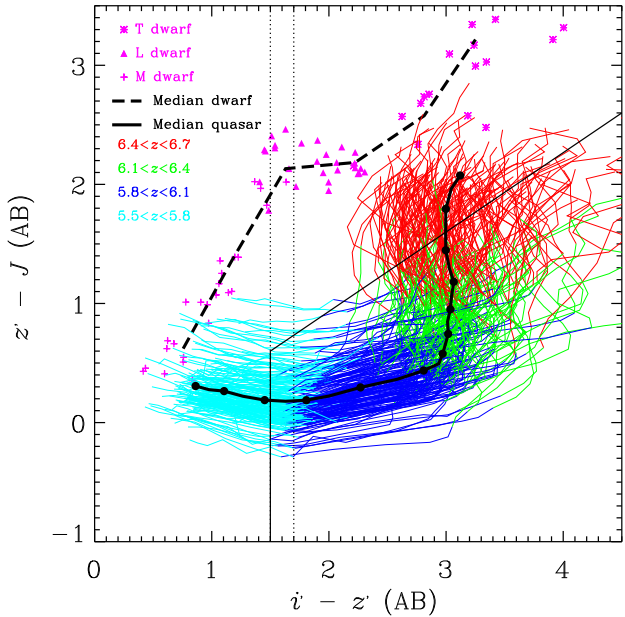}}
\caption{$i'-z'$ vs $z'-J$ in the CFHT filters for simulated quasars
and literature low mass stars. The coloured curves show the tracks in
colour-colour space of the 180 redshifted $z\sim3$ quasars as they are
moved from $z=5.5$ at lower-left to $z=6.7$ at upper-right
(Sec.\,\ref{simulq}). The median colours at each redshift are
calculated to produce the median quasar track (thick black
line). Intervals of $\delta z=0.1$ along this track are marked with
circles. The $i'-z'$ vs $z'-J$ colours of low mass stars from the
literature are shown as plus signs (M dwarfs), triangles (L dwarfs),
asterisks (T dwarfs) (Sec.\,\ref{simulbd}). The median star colours as
a function of type are shown with a thick dashed line. The thin solid
line denotes the high-redshift quasar selection region given in
eqns. 1 and 2. Dotted lines indicate $i'-z'=1.5$ and $i'-z'=1.7$.
\label{fig:colcolsimul}
}
\end{figure}

To determine the colours of quasars in our filters we convolve the 180
simulated quasar spectra with the $i', z'$ (MegaCam) and $J$ (CFHT-IR)
filters. The filter profiles include the effect of CCD quantum
efficiency and atmospheric transmission. All magnitudes are determined
on the AB system. Note this is different from the $z>5.7$ quasar
survey in the SDSS where $i'$ and $z'$ are on the AB system, but $J$
is on the Vega system\footnotemark. The colours are determined for a range of
redshifts from $z=5.5$ to $z=6.7$. We also ran the simulated quasar
spectra through the SDSS filter system to determine the SDSS colours
of our quasar templates. We found good agreement between the simulated
$i'-z'$ colours as a function of redshift and the observed colours of
the 12 $z>5.7$ SDSS quasars. This provides a check on our Lyman forest
absorption technique since the $i'-z'$ colours are particularly
sensitive to this absorption.

\footnotetext{$J_{\rm AB}=J_{\rm Vega}+0.90$ for the CFHT-IR filter.}

The $i'-z'$ vs $z'-J$ diagram is shown in Fig.\,\ref{fig:colcolsimul}. The colours of all
180 simulated quasars are plotted as a function of redshift from
$z=5.5$ at lower-left to $z=6.7$ at upper-right. Note the tracks of
the quasars in colour-colour space follow a similar trend to that
shown for a simulated quasar in the SDSS system in Fan et al. (2003).
However, there is a significant difference. The SDSS $i'$ and $z'$
filters are not overlapping, whereas there is a small, but
significant, overlap for the MegaCam filters. The effect of this
overlap is to decrease the $i'-z'$ colours of quasars, particularly in
the range $5.7<z<6$ where the \lya\ line passes through the overlap
region. Therefore we need to adopt a lower $i'-z'$ selection criteria
than the $i'-z'\geq 2.2$ used by the SDSS if we are to select quasars at
similar redshifts. The box bounded by the regions
\begin{equation}
\label{eqn:iz}
i'-z'\geq1.5
\end{equation}
\begin{equation}
i'-z'-1.5(z'-J)\geq0.6
\end{equation}
defines the default colour-colour high-redshift quasar selection
criteria. It can be seen from inspection of the figure that these
criteria select almost all quasars in the range $5.8<z<6.4$.

\begin{figure}
\resizebox{0.48\textwidth}{!}{\includegraphics{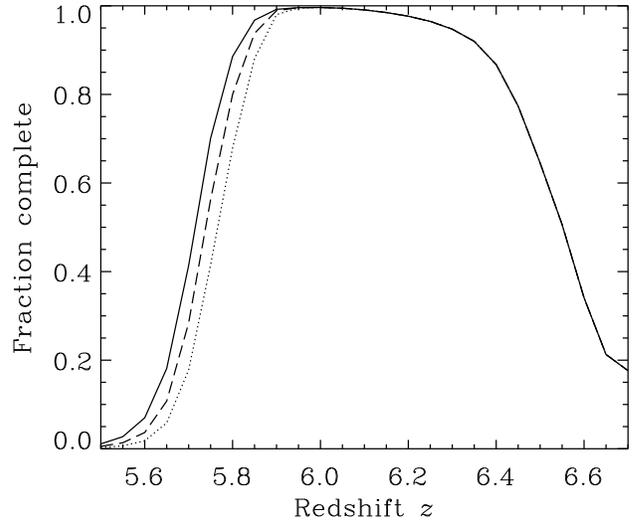}}
\caption{The fraction of quasars selected by the colour selection
criteria of eqns. 1 and 2 as a function of redshift (solid line). This
takes into account photometric scatter as described in
Sec.\,\ref{simulq}. The dashed and dotted lines show the fractions if
replacing eqn. 1 with $i'-z'\geq1.6$ and 1.7, respectively.
\label{fig:colcomp}
}
\end{figure}

To determine the quasar selection efficiency as a function of redshift
more quantitatively, we have determined the fraction of quasars as a
function of redshift which fall within this box. To account for the
fact that the colours of objects detected in CCD imaging have a
photometric uncertainty we model the location of each quasar at each
redshift in colour-colour space as a 2D gaussian probability density
distribution with $\sigma=0.2$. This error in colours is typical for
the faintest sources in our survey (Sec.\,\ref{nir}). We then
proceed to calculate the fraction of the probability that falls within
the selection region.  This process is repeated for three different
$i'-z'$ cuts of $1.5, 1.6$ and $1.7$ to determine the effect of
imposing different colour cuts on the selection efficiency as a
function of redshift.

The results of this process are shown in
Fig.\,\ref{fig:colcomp}. Quasars can be selected over the redshift
range $5.7<z<6.5$ at a completeness fraction of $>0.5$. The
completeness is close to unity at $5.8<z<6.4$.  The effect of imposing
a stricter cut of $i'-z'\geq1.6$ or 1.7 is fairly small and only has
an impact at the lower end of the redshift range (up to $z=5.9$). \\

\subsection{Ultracool stars}
\label{simulbd}

To determine the colours of ultracool stars (M, L and T dwarfs) in our
filters we convolve observed spectra obtained from Sandy Leggett's
online database\footnotemark\ with the $i'$, $z'$ (Mega Cam) and $J$
(CFHT-IR) filters. Only stars without significant gaps in spectral
coverage at the wavelengths of these filters were used. The spectra
are from the following sources: Tsvetanov et al. (2000), Burgasser et
al. (2000), Kirkpatrick et al. (2000), Reid et al. (2001), Leggett et
al. (2002) and Knapp et al. (2004).  Fig.\,\ref{fig:colcolsimul} shows
the dwarf colours as plus signs (M), triangles (L) and asterisks
(T). A sequence is also plotted by binning the dwarfs by type and
evaluating the median colour in each bin. This figure demonstrates
that the $i'-z'$ vs $z'-J$ diagram provides a very clean separation
between quasars and brown dwarfs for $i'-z'>$1.3. The $i'-z'\geq1.5$
selection criteria which we adopt nominally selects dwarfs cooler than
L2/L3. With significant noise in our colours and the much larger
population of late-M dwarfs in a flux-limited sample, a significant
number of slightly warmer stars may be {\it observed} to have
$i'-z'\geq1.5$.

\footnotetext{ftp://ftp.jach.hawaii.edu/pub/ukirt/skl/}

\section{Observations}

\subsection{CFHTLS Imaging}
\label{opt}

In this paper we use optical imaging from the Canada-France-Hawaii
Telescope Legacy Survey (CFHTLS)\footnotemark\ to search for
high-redshift quasars. The data used come from the first official
release: T0001.  This release includes data in the four Deep Survey
fields obtained during the period June 2003 - July 2004. The Deep
Survey has a 5-year lifetime, so the final dataset will go $\sim
1$\,mag deeper than the data used in this paper.

\footnotetext{http://www.cfht.hawaii.edu/Science/CFHTLS}

Full details of this dataset are available elsewhere\footnotemark, but
here we summarise the details relevant to our quasar search. Each of
the four Deep fields (named D1-4) is one dithered pointing with the
MegaCam camera.  Therefore the total area of each field is $\approx
1$\,sq.\,deg. The total integration time per field in the $i'$ and
$z'$ filters respectively are D1: 52.0\,ks, 12.2\,ks; D2: 18.5\,ks,
10.1\,ks; D3: 59.6\,ks, 15.1\,ks; D4: 58.8\,ks, 26.6\,ks. The seeing
FWHM is $\approx 0.9\asec$ in all the images. The astrometry of the
images in the two bands are aligned to better than a pixel ($0.186\asec$).

\footnotetext{http://terapix.iap.fr}

To find very red objects in this dataset we generated catalogues of
objects detected in the $z'$-band images using the Sextractor software
(Bertin \& Arnouts 1996). We ran Sextractor in ``double-image'' mode
to determine $i'$-band measurements for all the objects detected in
$z'$. Magnitudes in both images were measured in $2\asec$ circular
apertures. Aperture corrections to total magnitudes were applied for
unresolved sources. The magnitudes have been corrected for galactic extinction
using the maps of Schlegel, Finkbeiner \& Davis (1998). Note that
stars may not lie behind the total galactic obscuring column, but
since the main goal of this paper is quasar selection and the applied
galactic extinction corrections are smaller than our photometric
errors, we do not attempt any further corrections for stars.

From these catalogues we drew a sample of objects satisfying the
colour criterion of eqn. \ref{eqn:iz}. We restricted the selection to
sources which have $z'$-band photometric errors $\sigma(z')\leq 0.10$
to ensure we do not spend time following up spurious
detections. Although images of these fields in the $u^{*}, g'$ and
$r'$ filters are also part of the T0001 release, we did not impose any
other colour criteria since the $i'$ images have the longest
integration time so most selected objects are undetected at shorter
wavelengths.  Every candidate was inspected by eye to ensure that it
is real. There were many spurious sources, mostly located near to very
bright stars, where the diffraction patterns in the $i'$ and $z'$
images were different.  After excising these from our target list we
are left with 7, 12, 11 and 5 objects satisfying eqn. \ref{eqn:iz} and
$\sigma(z')\leq 0.10$ in fields D1, D2, D3 and D4, respectively.

\subsection{Near-infrared imaging}
\label{nir}

As shown for the SDSS by Fan et al. (2001) and discussed in
Sec.\,\ref{simul}, the most efficient way of discriminating between
high-redshift quasars and brown dwarfs is by using near-infrared
$J$-band photometry. Brown dwarfs and very low mass stars have red
$z'-J$ colours and quasars have blue $z'-J$ colours. The exact colour
selection criteria used in the CFHQS to separate stars and quasars was
given in Sec.\,\ref{simulq}.

\begin{figure}
\resizebox{0.48\textwidth}{!}{\includegraphics{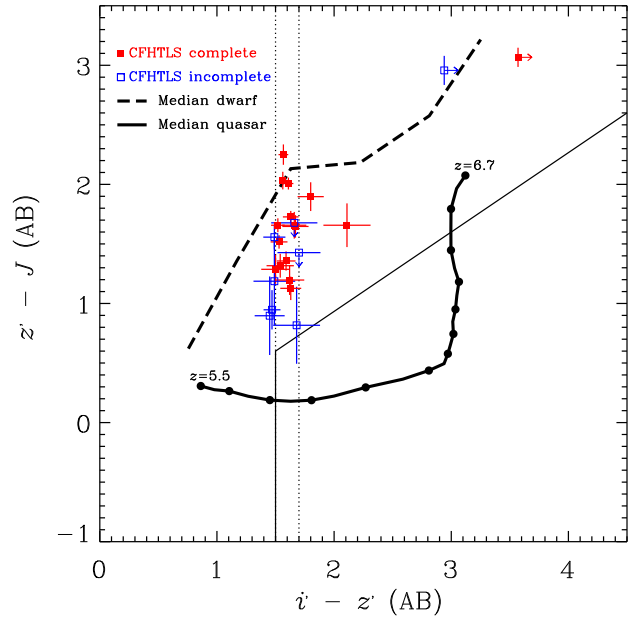}}
\caption{Very red objects selected from the CFHTLS Deep $i'$ and $z'$
imaging which were observed with CFHT-IR at $J$-band are shown by
squares with $1\sigma$ error bars. Filled squares show those objects
belonging to the complete sample of candidates defined by
$i'-z'\geq1.5$, $\sigma(z') \leq 0.10$ and $z'< 23.35$. Open squares
show sources observed which fail one or more of these criteria. Also
shown are the median colours of simulated quasars (thick solid line)
and low mass stars (thick dashed line) from
Fig.\,\ref{fig:colcolsimul}.  The thin solid line denotes the
high-redshift quasar selection region given in eqns. 1 and 2. Dotted
lines indicate $i'-z'=1.5$ and $i'-z'=1.7$.
\label{fig:colcoldata}
}
\end{figure}

\begin{deluxetable*}{cccccc}
\tabletypesize{\scriptsize}
\tablecaption{Table of observed objects. \label{tbl-1}}
\tablewidth{0pt}
\tablehead{
\colhead{Field} &   \colhead{Name}   & \colhead{$i'$ (AB)} &
\colhead{$z'$ (AB)} & \colhead{$J$ (AB)}  & \colhead{Complete?}  } 
\startdata
D1 &  CFHTLS\,J022506.79-042958.4 & $25.41 \pm 0.18$ &  $23.30 \pm 0.09$ &  $21.64 \pm  0.16$ &  $\checkmark$  \\
D1 &  CFHTLS\,J022456.72-045905.9 & $24.30 \pm 0.10$ &  $22.80 \pm 0.07$ &  $21.51 \pm  0.10$ &  $\checkmark$  \\
D1 &  CFHTLS\,J022714.12-045723.1 & $24.35 \pm 0.10$ &  $22.81 \pm 0.06$ &  $21.49 \pm  0.08$ &  $\checkmark$  \\
D1 &  CFHTLS\,J022455.66-040602.2 & $21.78 \pm 0.03$ &  $20.17 \pm 0.03$ &  $18.16 \pm  0.04$ &  $\checkmark$  \\
D1 &  CFHTLS\,J022415.59-040431.1 & $23.80 \pm 0.07$ &  $22.21 \pm 0.04$ &  $20.85 \pm  0.07$ &  $\checkmark$  \\
D1 &  CFHTLS\,J022559.39-040400.0 & $23.47 \pm 0.06$ &  $21.84 \pm 0.03$ &  $20.11 \pm  0.04$ &  $\checkmark$  \\
D1 &  CFHTLS\,J022728.09-040318.7 & $23.49 \pm 0.06$ &  $21.96 \pm 0.04$ &  $20.44 \pm  0.04$ &  $\checkmark$  \\
D2 &  CFHTLS\,J100113.06+022622.4 & $>26.35$         &  $22.78 \pm 0.07$ &  $19.71 \pm  0.04$ &  $\checkmark$  \\
D2 &  CFHTLS\,J100142.08+014931.8 & $24.22 \pm 0.10$ &  $22.55 \pm 0.05$ &  $20.90 \pm  0.08$ &  $\checkmark$  \\
D2 &  CFHTLS\,J100130.42+015952.8 & $24.10 \pm 0.09$ &  $22.44 \pm 0.05$ &  $20.78 \pm  0.10$ &  $\checkmark$  \\
D2 &  CFHTLS\,J095928.34+021819.8 & $24.37 \pm 0.11$ &  $22.75 \pm 0.06$ &  $21.55 \pm  0.11$ &  $\checkmark$  \\
D2 &  CFHTLS\,J095922.77+020207.9 & $23.71 \pm 0.08$ &  $22.08 \pm 0.04$ &  $20.95 \pm  0.09$ &  $\checkmark$  \\
D2 &  CFHTLS\,J100132.04+020022.4 & $22.91 \pm 0.05$ &  $21.39 \pm 0.03$ &  $19.73 \pm  0.05$ &  $\checkmark$  \\
D3 &  CFHTLS\,J141659.91+521521.9 & $23.89 \pm 0.04$ &  $22.32 \pm 0.03$ &  $20.07 \pm  0.08$ &  $\checkmark$  \\
D3 &  CFHTLS\,J141731.77+523920.4 & $23.73 \pm 0.04$ &  $22.17 \pm 0.03$ &  $20.13 \pm  0.07$ &  $\checkmark$  \\
D4 &  CFHTLS\,J221705.05-172932.5 & $24.47 \pm 0.10$ &  $22.67 \pm 0.05$ &  $20.77 \pm  0.11$ &  $\checkmark$  \\
D1 &  CFHTLS\,J022757.64-043019.5 & $25.03 \pm 0.16$ &  $23.35 \pm 0.12$ &  $22.53 \pm  0.30$ &  $\times$   \\
D1 &  CFHTLS\,J022516.57-041402.9 & $24.87 \pm 0.14$ &  $23.38 \pm 0.11$ &  $22.19 \pm  0.35$ &  $\times$   \\
D1 &  CFHTLS\,J022614.42-044600.4 & $24.26 \pm 0.10$ &  $22.81 \pm 0.08$ &  $21.91 \pm  0.32$ &  $\times$   \\
D1 &  CFHTLS\,J022734.82-042438.7 & $23.52 \pm 0.06$ &  $22.05 \pm 0.04$ &  $21.10 \pm  0.16$ &  $\times$   \\
D2 &  CFHTLS\,J095914.80+023655.2 & $>26.35$         &  $23.41 \pm 0.10$ &  $20.45 \pm  0.07$ &  $\times$   \\
D4 &  CFHTLS\,J221430.53-180230.5 & $25.11 \pm 0.16$ &  $23.41 \pm 0.09$ &  $>21.98$          &  $\times$   \\
D4 &  CFHTLS\,J221457.74-172104.2 & $25.32 \pm 0.17$ &  $23.66 \pm 0.10$ &  $>21.98$          &  $\times$   \\
D4 &  CFHTLS\,J221437.22-180417.4 & $24.08 \pm 0.08$ &  $22.59 \pm 0.05$ &  $21.03 \pm  0.09$ &  ~$\times$
\enddata
\tablecomments{Objects selected by their red $i'-z'$ colours from the
CFHTLS Deep fields which were observed at $J$-band with CFHT-IR. The
magnitudes listed are total magnitudes derived from aperture
magnitudes. The final column indicates whether the object satisfies
the $i'$ and $z'$ selection criteria which form the basis for our
complete sample of quasar candidates for $J$-band imaging. These
criteria are $i'-z'\geq1.5$, $\sigma(z') \leq 0.10$ and $z'<
23.35$. The three-colour photometry indicates that all these objects
are ultracool dwarfs (Fig.\,\ref{fig:colcoldata}). Their properties
are discussed further in Sec.\,\ref{bd}.}
\end{deluxetable*}

Observations at $J$-band were performed using the CFHT-IR camera at
the CFHT. CFHT-IR is a HgCdTe $1024^2$ array with pixel size $0.211\,
\asec$. Observations were carried out in photometric conditions on UT
2003 November 9, 10, 2004 May 4 and 2004 November 23, 24. Our first
observations with CFHT-IR came a year before the T0001 release of the
CFHTLS Deep data. For the first two observing runs we used preliminary
reductions of the CFHTLS dataset generated by one of us (SDJG) and
also by David Balam and Chris Pritchet. This led to the observation of
some objects which now have $i'-z'<1.5$ in the most recent (and
deepest) reduction.  Due to constraints imposed by scheduling and poor
weather it has not been possible to observe a complete sample of
$i'-z'>1.5$ sources to a consistent magnitude limit in all 4
fields. We prioritised the reddest and brightest sources so that we
can quantify our incompleteness.  The sources not observed all lie
close to the $i'-z'=1.5$ boundary and therefore we can account for the
possibility they may be quasars at $z\approx 5.8$ in our selection
function (Sec.\,\ref{complete}).

A total of 24 sources were observed at $J$-band of which 20 have
$i'-z'\geq1.5$ (the rest have $1.45 \leq i'-z'<1.5$). 15 out of the 35
sources identified in Sec.\,\ref{opt} were not observed (five in D2,
nine in D3 and one in D4). The poor completeness in D3 is a
consequence of bad weather during the CFHT-IR run in May 2004.
$J$-band magnitudes on the AB scale were measured using $2\asec$
circular apertures and aperture corrections applied to give total
magnitudes.

The resulting photometry is given in Table\,1. The $i'-z'$ vs $z'-J$
colours of these objects are plotted in Fig.\,\ref{fig:colcoldata}. As
is immediately clear in this diagram, the colours of all 24 targeted
objects are consistent with stars and inconsistent with quasars. The
three closest to the quasar selection region with $z'-J<1$ are not
part of the complete sample. Two of these have $i'-z'<1.5$. The other
has $\sigma(z')> 0.10$ and a large uncertainty on its $J$-band
magnitude and hence $z'-J$ colour. It is possible that this object is
a quasar, but given the poor constraint on its colour and the fact
that stars are many times more common than quasars at this $i'-z'$
colour, the prior on the likelihood is that it is a star.

The two reddest objects (at $z'-J \approx 3$) do lie close to the
expected colours of $z>6.8$ quasars. However such a high redshift
would require very high-luminosity quasars since at this redshift the
Lyman-$\alpha$ line is moving out of the $z'$-band. The luminosity of
such quasars would be comparable to those in the SDSS at $z\sim 6$
which have a surface density of $\approx 1$ per 500 square degrees
(Fan et al. 2004). The probability of finding such a quasar in 3.8
square degrees is very low. There is a much greater probability that
they are T dwarfs (see Sec.\,\ref{bd}). Even if these objects were
quasars it would not affect the results of the rest of this paper
because we only use the survey volume for quasars out to $z=6.4$.

\subsection{Ultracool dwarfs detected} 
\label{bd}

The two reddest objects (CFHTLS\,J095914.80+023655.2, $i'-z'>2.9$ and
CFHTLS J100113.06+022622.4, $i'-z'>3.5$) both have $z'-J \approx 3$
and are T dwarfs. The constraint from the $i'-z'$ and $z'-J$ colours
indicate that the former is cooler than T1 and the latter cooler than
T4. To characterise their type more effectively we obtained $H$-band
photometry of them with CFHT-IR. Both objects are very blue at $J-H$
confirming their T dwarf nature (even bluer than a $z\sim7$
quasar). CFHTLS\,J095914.80+023655.2 has $J-H({\rm AB})=-0.26 \pm
0.09$ and CFHTLS\,J100113.06+022622.4 has $J-H({\rm AB})=-0.70 \pm
0.08$.  Using the transformation $J-H({\rm Vega})=J-H({\rm AB})+0.44$
for comparison with the plot of $J-H$ vs spectral type in Leggett et
al. (2002), we find that CFHTLS\,J095914.80+023655.2 is type T3 and
CFHTLS\,J100113.06+022622.4 is type T6. With J$_{\rm AB} \approx 20$,
they are $3-4$ magnitudes fainter than most SDSS and 2MASS T
dwarfs. Approximate distances based upon the relationship between
absolute magnitude and spectral type (Tinney, Burgasser and
Kirkpatrick 2003) are 150 and 70\,pc, respectively, making them two of
the most distant T dwarfs known.

Only four or five of the analysed objects have colours typical of L dwarfs
($i'-z'>1.5$ and $z'-J \sim 2.0$). The $z'-J<1.7$ colours of the rest
indicate that they are late-M dwarfs, with a true $i'-z'$ below 1.5,
scattered into our $i'-z'$ box by 1-3\,$\sigma$ random noise excursions
(this scattering is also apparent in the SDSS sample, see e.g.,
Fig.\,3 of Fan et al. 2003). Within the limited statistics, the 
number of detected L dwarfs is consistent with an extrapolation of
the densities seen in the shallow whole sky surveys.

\subsection{Completeness}
\label{complete}

To turn the non-detection of quasars in this dataset into a useful
limit on the quasar space-density and luminosity function, it is
necessary to evaluate the completeness of our survey. There are two
relevant selection criteria which will affect the completeness as a
function of redshift and absolute magnitude:
\begin{itemize}
\item{The $z'$-band magnitude limit as a function of the survey area}
\item{The colour selection criteria}
\end{itemize}

Excluding the field edges which we have masked out of our
catalogues, the $z'$-band images show a fairly uniform noise and
therefore detection limit across the field. Since the majority of our
candidates have $z' < 23.35$ and point-sources of this magnitude have
$\sigma(z') \leq 0.10$\, across most of the survey, we adopt a magnitude limit
for the present analysis of $z'< 23.35$. To determine the efficiency
with which objects could be found as a function of magnitude we ran
simulations by inserting 10\,000 artificial stars into each $z'$-band
image and using Sextractor to recover them. Even at bright magnitudes,
about 5\% of sources are not recovered because they overlap with other
objects or regions affected by bright stars. We note that this
fraction is much lower than the $\approx 20\%$ of the area in the mask
region file released by Terapix. We find their masking to be much too
conservative for our purpose. By $z'=23.5$ the recovered fraction has
decreased to $\approx 80\%$ in all the images. For each field we
determine the area in which sources can be found as a function of
their magnitude. At bright magnitudes the total area surveyed is 3.83
sq deg, dropping to 3.32 sq deg. at our limit of $z'=23.35$.

The colour selection criteria were discussed extensively with
simulated quasar and stellar colours in Sec.\,\ref{simulq}. Here we
recap these selection criteria and evaluate the extent to which colour
selection impacts our completeness. The nominal colour selection
criteria are $i'-z'\geq1.5$ and $i'-z'-1.5(z'-J)\geq0.6$. No
observed objects lie within this box. Since we were unable to observe
all the candidates with $i'-z'>1.5$ down to the magnitude limit, our
completeness level in terms of the $i'-z'$ colour varies from
field-to-field. Field D1 is complete to $i'-z'\geq1.5$, D2 and D3 are
complete to $i'-z'\geq1.7$ and D4 is complete to $i'-z'\geq1.6$. Note
that even though only two out of eleven candidates in D3 were
observed, all the unobserved objects have $1.5\leq i'-z'<1.7$.

Fig.\,\ref{fig:colcomp} showed the completeness due to colour cuts as determined from
the simulated high-redshift quasar colours. The effect of implementing
redder cuts than the nominal $i'-z'\geq1.5$ is to exclude quasars in
the range $5.7<z<5.9$. The completeness at $z=5.8$ is $89\%, 80\%$ and
$68\%$ for $i'-z'\geq1.5, 1.6$ and $1.7$, respectively. At $z=5.9$,
the colour completeness for all three cuts are $>98\%$. Therefore using
a redder cut in $i'-z'$ makes only a very small difference to our
sensitivity to quasars across the full range of $5.8<z<6.4$. This
means that we can include fields like D2 and D3 which have not been so
well followed up in the near-infrared whilst losing only a small
amount in the completeness of high-redshift quasar selection.

The magnitude and colour completenesses are combined to generate an
effective volume of our high-redshift quasar survey as a function of
redshift (over the interval $5.8<z<6.4$) and absolute
magnitude. Absolute magnitudes of the quasar continuum at a rest-frame
wavelength of 1450\AA\ are calculated for the possible range of
apparent $z'$-band magnitudes. These calculations use $k$-corrections
generated from the mean $k$-corrections of the 180 redshifted
$z\approx 3$ quasars discussed in Sec.\,\ref{simulq}.

\section{Constraining the quasar luminosity function}
\label{lf}

We now use the non-detection of high-redshift quasars in the CFHTLS
Deep fields to constrain the luminosity function (LF).  We parametrise
the LF using the double power-law form that provides a good fit at
lower redshift (Croom et al. 2004), i.e.
\begin{equation}
\Phi (M_{1450},z) = \frac{\Phi(M_{1450}^{*})}{10^{0.4(\alpha+1)(M_{1450}-
M_{1450}^*)} + 10^{0.4(\beta+1)(M_{1450}-M_{1450}^*)}}.
\end{equation}
The luminosity distribution of SDSS $z\approx6$ quasars from Fan et
al. (2004) is used to fit the bright-end normalisation
$\Phi(M_{1450}=-26)$ as a function of the bright-end slope
$\beta$. Taking the range in $\beta$ allowed by the SDSS and the value
of $\Phi(M_{1450}=-26)$ at each $\beta$, we set up an array of
possible double-power-law LFs. The faint-end slope $\alpha$ will not
be strongly constrained by our data and therefore we do not treat this
as a free parameter. Instead we assume that $\alpha$ takes a value of
$-1.0, -1.5$ or $-2.0$ and repeat our analysis for each of these
values. The best-fit values at the highest redshifts at which $\alpha$
has been measured ($z\sim 2-3$) lie in the range $-1.1$ to $-1.8$
(Boyle et al. 2000; Croom et al. 2004; Hunt et al. 2004; Richards et
al. 2005). We will consider the effect of $\alpha$ later
on. Sec.\,\ref{complete} discussed our survey volume as a function of
redshift and absolute magnitude. We use the array of possible LFs
combined with the survey volume to calculate the number of $5.8<z<6.4$
quasars $\mu$ that we would expect to have discovered in our survey as
a function of $\beta$ and break magnitude $M^*_{1450}$. The
probability of not finding any quasars given this LF is then simply
given by the Poisson probability $P= e^{-\mu}$.

\begin{figure}
\resizebox{0.48\textwidth}{!}{\includegraphics{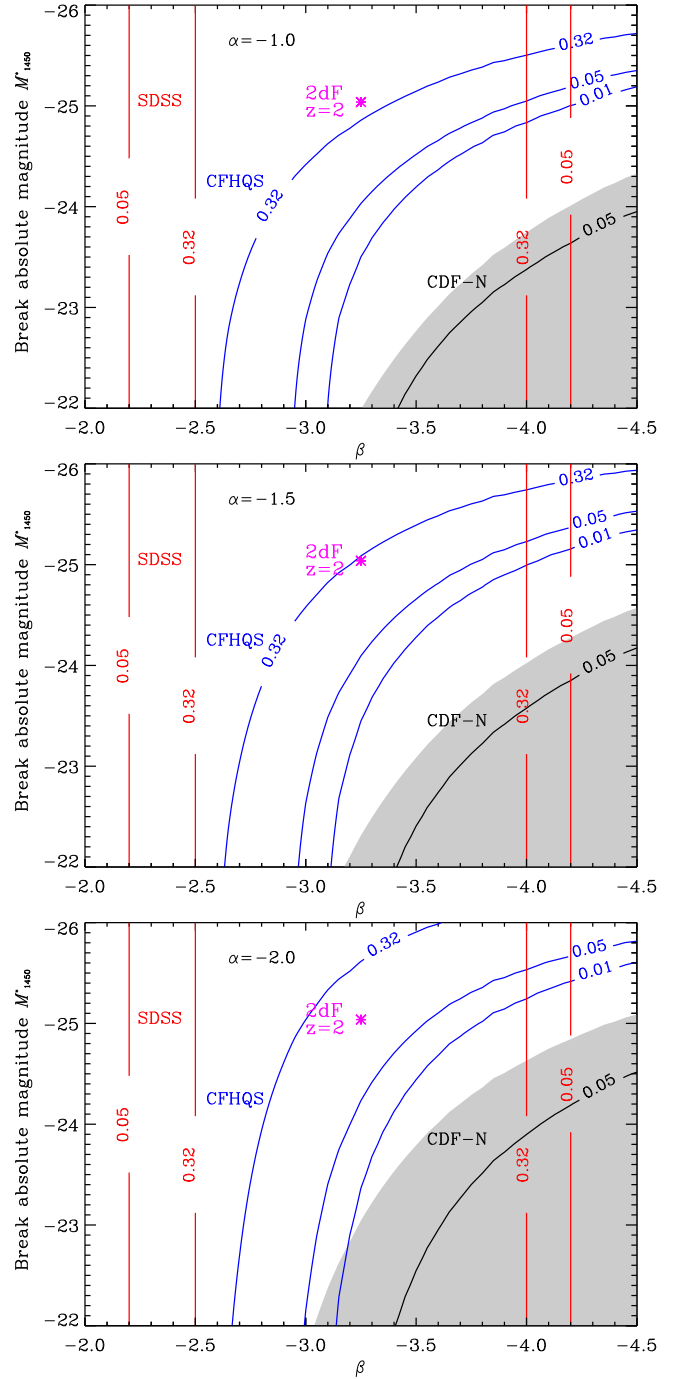}}
\caption{Constraints on the $z=6$ quasar luminosity function from a
variety of optical surveys. Each panel shows the constraints on the
bright-end slope $\beta$ and break magnitude $M^*_{1450}$. The three
panels assume values for the faint-end slope of $\alpha=-1$ (upper),
$-1.5$ (middle) and $-2$ (lower). Confidence intervals are drawn for
the CFHQS, SDSS and the CDF-N (Sec.\,\ref{lf}).  The best fit
parameter values for the 2dF at $z=2$ (Croom et al. 2004) are shown
for reference. The grey shaded region shows the values of parameters
necessary for the $z=6$ quasar population to produce enough ionizing
photons to reionize the universe as described in Sec\,\ref{ion}.
\label{fig:lfparam}
}
\end{figure}

\begin{figure}
\resizebox{0.48\textwidth}{!}{\includegraphics{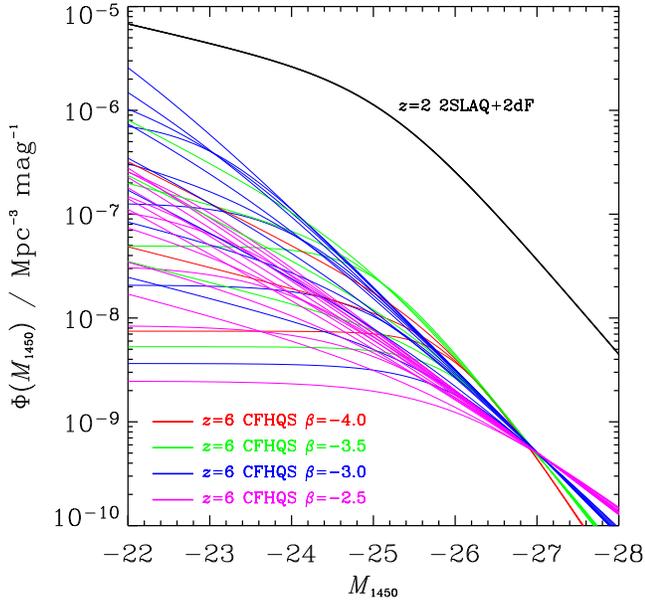}}
\caption{Plausible double power-law parametrisations of the $z=6$
quasar optical luminosity function are shown as colored lines. Only
those models with confidence levels $\geq 0.05$ based on the CFHQS and
SDSS constraints are shown. These models are normalized to the SDSS
results of Fan et al. (2004) which is why they have approximately the
same density at $M_{1450}=-27$. The luminosity function at $z=2$ from
the 2SLAQ\,+\,2dF surveys (Richards et al. 2005) is shown as a black
line. Our work finds that the large decrease in the space density of
luminous quasars from $z=2$ to $z=6$ found by the SDSS also extends to
lower luminosities ($M_{1450}\sim-24$).
\label{fig:lf}
}
\end{figure}

The result of this analysis is shown in Fig.\,\ref{fig:lfparam}.
Curves are plotted at confidence levels of $0.32, 0.05$ and
$0.01$. Separate panels show the different results obtained by
assuming faint-end slopes of $\alpha=-1, -1.5, -2$. Since our survey
is only sensitive to quasars brighter than $M_{1450}=-23$, our
constraints depend weakly on $\alpha$ unless the break magnitude is
very bright ($M^*_{1450}<-25$).  The evolution in the 2dF quasar LF is
fit by pure-luminosity evolution with the break evolving from
$M^*_{1450}=-21$ at $z=0$ to $M^*_{1450}=-25$ at $z=2$ (Croom et
al. 2004). Since the number density of luminous optically-selected
quasars declines at redshifts beyond $2-3$ (Fan et al. 2004),
continued pure-luminosity evolution would lead to a fainter
$M^*_{1450}$ at $z=6$ than at $z=2$. Meiksin (2005) determined that a
model with continued pure-luminosity evolution that fits the luminous
SDSS $z=6$ sample would have $M^*_{1450}=-22.9$ at $z=6.0$. Although
we do not necessarily expect the high-redshift evolution to be
characterised by luminosity evolution, it is unlikely that the break
magnitude would be brighter than it is at $z=2$, so we consider
$M^*_{1450}=-25$ as a likely lower limit.

Barger et al. (2003) reported a negative search for $z>5.2$ quasars in
the Chandra Deep Field-North (CDF-N). Their work goes deeper than our
study (to $z'=25.2$) but covers a much smaller area (0.031
sq. deg). We have therefore repeated our analysis of LF constraints
using the lack of $5.8<z<6.4$ quasars in the CDF-N.  The 0.05
confidence contour on $\beta$ and $M^*_{1450}$ from the CDF-N is
plotted on Fig.\,\ref{fig:lfparam}. It can be seen that our constraint
from CFHQS is considerably stronger than that from the CDF-N due to
our survey volume being $\approx 100$ times greater. Also plotted on
Fig.\,\ref{fig:lfparam} are the constraints on $\beta$ from the
luminosity distribution of the SDSS (Fan et al. 2004). The CFHQS
provides much better lower limits on $\beta$ than the SDSS. For all
the values of $\alpha$ considered, the 0.05 confidence limit is
$\beta>-3.2$ if $M^*_{1450}=-24$ and $\beta>-3.1$ if $M^*_{1450}=-23$.

Fig.\,\ref{fig:lf} shows the LF constraints from the CFHQS more
directly. The colored curves correspond to LF models which are allowed
at $\geq 0.05$ confidence. We can use the constrained LF parameters to
place an upper limit on the space-density of quasars at $z=6$ brighter
than a certain absolute magnitude limit. We adopt a limit of
$M_{1450}<-23.5$ since at this magnitude cut there is almost no
dependence of the density upper limit on $\alpha, \beta$, or
$M^*_{1450}$. We find that
$\rho(M_{1450}<-23.5)<1.7\times10^{-7}\,{\rm Mpc}^{-3}$ at 95\%
confidence. This limit is a factor of 25 lower than the density at
$z=2$ measured in the 2SLAQ\,+\,2dF surveys (Richards et al. 2005).
This confirms that the decline in the space-density of luminous
quasars at $z\sim6$ (Fan et al. 2004) is mirrored by the lower
luminosity population. A decline in the space-density of low
luminosity quasars has been seen in the range $z\sim3-5$ (Barger et
al. 2003; Wolf et al. 2003; Cristiani et al. 2004; Sharp et al. 2004),
but the large volume and faint magnitude limit of our survey extends
this result to $z\sim6$.

\section{A limit on the quasar contribution to the ionizing background at $z\sim6$}
\label{ion}

If the high-redshift quasar LF can be fit by a double power-law with a
break at $M^*_{1450} \ltsimeq -23$ and $\alpha>-2$, then a large
fraction of the luminosity emitted by the quasar population will be
emitted by quasars that we could have found in our survey. This means
that our survey is useful for providing an upper limit to the number
density of ionizing photons emitted by quasars at $z=6$.

To estimate the number of hydrogen-ionizing photons emitted by a quasar with a given $M_{1450}$, we assume the
typical spectral shape determined for a large sample of quasars by
Telfer et al. (2002). We assume that 100\% of these photons escape
from the host galaxy.  The ionizing photon emission density is
calculated by multiplying the photon emission rate per quasar by the
number density of quasars and integrating over the range of quasar
luminosities. We perform the integration down to quasars as faint as
$M_{1450}=-18$.

\begin{figure}
\resizebox{0.48\textwidth}{!}{\includegraphics{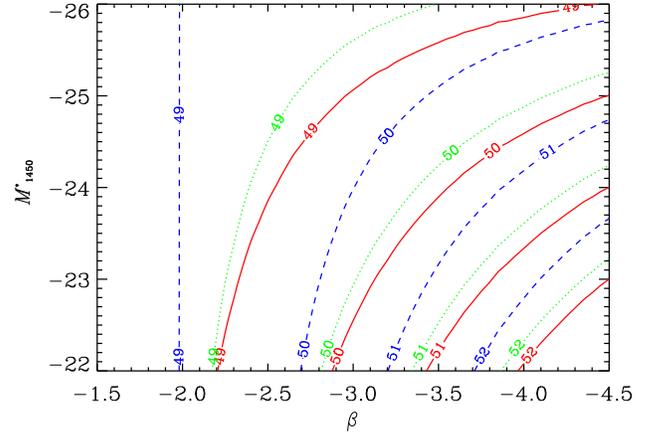}}
\caption{The ionizing photon production rate density for the $z=6$
quasar population as a function of $\beta$ and $M^*_{1450}$. Contours
are drawn for three values of the luminosity function faint-end slope:
$\alpha=-1$ (solid), $-1.5$ (dotted) and $-2$ (dashed). Contours are
plotted at $10^{49}, 10^{50}, 10^{51},
10^{52}$\,photons\,s$^{-1}$\,Mpc$^{-3}$.
\label{fig:ion}
}
\end{figure}

Fig.\,\ref{fig:ion} shows the ionizing photon production rate density
for the $z=6$ quasar population as a function of $\beta$ and
$M^*_{1450}$.  Contours are plotted for the three different values of
$\alpha$ considered in the previous section. Note that the shape of
the contours is quite similar to the shape of the confidence limits on
these parameters determined from our survey for high-redshift
quasars. This is due to the reason outlined previously that the main
constraint from our survey is on the number density of quasars
brighter than $M_{1450}=-23$ and such quasars dominate the ionizing
photon budget unless $\alpha$ is very steep. We note that the ionizing
photon production rate density in Fig.\,\ref{fig:ion} is consistent
with the calculations of Fan et al. (2001), but not with those of Yan
\& Windhorst (2004) who found a factor of 10 lower rate for the same
quasar LF parameters. The reason for this discrepancy is unclear.

There is considerable debate about the photon density required to
reionize the diffuse neutral hydrogen in the universe at $z=6$. The
main uncertainty is due to recombination of ionized hydrogen atoms,
which depends upon the clumpiness of the IGM. A detailed discussion of
the clumpiness issue is beyond the scope of this paper. We follow
Meiksin (2005) and assume that the clumpiness leads to 5 ionizing
photons per hydrogen atom necessary for reionization at $z=6$. Note
that this is substantially fewer than the 30 ionizing photons per
hydrogen atom used by Madau et al. (1999) which has been widely used
by other authors. Under our assumptions the ionizing photon production
rate density necessary for reionizing the universe at $z=6$ is
$3\times10^{50}$\,photons\,s$^{-1}$\,Mpc$^{-3}$.

The region of $\beta-M^*_{1450}$ space where the quasar population
alone can provide the necessary ionizing flux is shown by the grey
shaded regions in Fig.\,\ref{fig:lfparam}. For any value of $\alpha$ the
shaded region does not overlap with the CFHQS 0.05 confidence contours
and quasars do not provide enough ionizing photons to reionize the
universe. For $\alpha<-2$, the quasar emission is dominated by low
luminosity quasars and a population with a break at $M^*_{1450}>-23$
could still reionize the universe, although the details then depend
critically upon how far down the LF one integrates. All the evidence
at intermediate redshifts points towards $\alpha\geq -1.8$ (Boyle et al. 2000;
Croom et al. 2004; Hunt et al. 2004; Barger et al. 2005; Richards et
al. 2005; but see Hao et al. 2005 for evidence that $\alpha \approx -2$ in
the local universe) making it unlikely that $\alpha$ could be steeper than
$-2$ at $z=6$. We conclude that a 95\% confidence upper limit on the
ionizing photon rate density provided by quasars at $z=6$ is
$3\times10^{50}$\,photons\,s$^{-1}$\,Mpc$^{-3}$.

\section{Conclusions}

We have used 3.83 sq. deg. of optical imaging with a magnitude limit
of $z'=23.35$ to search for high-redshift quasars. Near-infrared
follow-up of candidates selected by their red $i'-z'$ colours all have
$z'-J$ colours consistent with being low mass stars: two are T dwarfs,
four or five are L dwarfs, and the rest are late-M dwarfs scattered
into our selection box by photometric errors. 

The lack of quasars at $z>5.8$ in these data has been used to provide
new constraints on the luminosity function. Several surveys underway
at CFHT will image between $100-1000$ sq. deg. in multiple filters to
somewhat shallower depths than the data in this paper. Assuming the
bright-end slope is not much flatter than the value at $z=4$ of
$\beta=-2.6$ (Fan et al. 2001), these surveys will allow the
identification of many $z>5.8$ quasars giving a much more accurate
determination of the luminosity function and provide targets for
studying the ionization state of the IGM.

We have considered the ionizing photon output at $z=6$ and concluded
that quasars provide $< 3 \times
10^{50}$\,photons\,s$^{-1}$\,Mpc$^{-3}$ at 95\% confidence and
assuming that the faint end slope $\alpha>-2$. The ionizing photon
output from the $z\sim6$ star-forming galaxy population is subject to
significant uncertainties as outlined earlier, but the best estimates
give a rate $\gtsimeq 10^{51}$\,photons\,s$^{-1}$\,Mpc$^{-3}$ (Yan \&
Windhorst 2004; Bunker et al 2004; Stiavelli et al. 2004). Therefore
quasars provide less than 30\% as many ionizing photons as galaxies at
$z\sim6$.  The evolution of the neutral hydrogen fraction inferred
from quasar absorption spectra shows that the universe had been almost
completely reionized by $z=6$ (Fan et al. 2004; Songaila 2004). Since
something has to keep the universe reionized at this redshift, the
most straightforward explanation is that the galaxy population is
responsible.

\acknowledgments Thanks to Jacqueline Bergeron, Kuenley Chiu, David
Crampton, Xiaohui Fan, John Hutchings, Alain Omont, Marcin Sawicki,
David Schade and Luc Simard for interesting discussions. Thanks to
David Balam and Chris Pritchet for providing their early reduction of
the CFHTLS imaging and an anonymous referee for useful comments. Based on
observations obtained with MegaPrime/MegaCam, a joint project of CFHT
and CEA/DAPNIA, at the Canada-France-Hawaii Telescope (CFHT) which is
operated by the National Research Council (NRC) of Canada, the
Institut National des Science de l'Univers of the Centre National de
la Recherche Scientifique (CNRS) of France, and the University of
Hawaii. This work is based in part on data products produced at
TERAPIX and the Canadian Astronomy Data Centre as part of the
Canada-France-Hawaii Telescope Legacy Survey, a collaborative project
of NRC and CNRS. Funding for the creation and distribution of the SDSS
Archive has been provided by the Alfred P. Sloan Foundation, the
Participating Institutions, the National Aeronautics and Space
Administration, the National Science Foundation, the U.S. Department
of Energy, the Japanese Monbukagakusho and the Max Planck Society. The
SDSS Web site is http://www.sdss.org/.

\end{document}